\documentclass[aps,plb,twocolumn,showpacs,superscriptaddress,groupedaddress]{revtex4}

\usepackage{graphicx}
\usepackage{dcolumn}
\usepackage{bm}
\usepackage{amssymb,amsmath}  
\usepackage{epsfig}
\usepackage{psfrag}
\usepackage{xspace}
\usepackage{multirow}
\usepackage{lscape}
\usepackage{comment}

\newcommand{\sigmatt}{\mbox{${\sigma}_{t\bar{t}}$}\xspace}

\newcommand{\dzero}     {D0}
\newcommand{\ttbar}     {\mbox{$t\bar{t}$}\xspace}

\newcommand{\ppbar}     {\mbox{$p\bar{p}$}\xspace}

\newcommand{\pythia}    {\mbox{\textsc{pythia}}}
\newcommand{\alpgen}    {\mbox{\textsc{alpgen}}}
\newcommand{\mcatnlo}    {\mbox{\textsc{mc@nlo}}}
\newcommand{\geant}     {\mbox{\textsc{geant}}}

\newcommand{\met}{\mbox{\ensuremath{E\kern-0.6em\slash_T}}\xspace}
\newcommand{\metx}{\mbox{\ensuremath{E\kern-0.6em\slash_x}}\xspace}
\newcommand{\mety}{\mbox{\ensuremath{E\kern-0.6em\slash_y}}\xspace}
\newcommand{\sigmet}{\ensuremath{\sigma_{E \kern-0.4em\slash_T}}\xspace}

\newcommand{\ttfull}{\mbox{$t\bar{t} \rightarrow W^{+}b \,
    W^{-}\bar{b} \rightarrow \ell^{+} \nu b \, \ell^{-} \bar{\nu} \bar{b}$}}

\newcommand{\herwig}    {\textsc{herwig}\xspace}

\newcommand{\ttb}{\mbox{$t\bar{t}$}\xspace}

\begin{document}

\hspace{5.2in} \mbox{FERMILAB-PUB-11/196-E}

\title{Measurement of spin correlation in \boldmath $t\bar{t}$ production using
  a matrix element approach}

\affiliation{Universidad de Buenos Aires, Buenos Aires, Argentina}
\affiliation{LAFEX, Centro Brasileiro de Pesquisas F{\'\i}sicas, Rio de Janeiro, Brazil}
\affiliation{Universidade do Estado do Rio de Janeiro, Rio de Janeiro, Brazil}
\affiliation{Universidade Federal do ABC, Santo Andr\'e, Brazil}
\affiliation{Instituto de F\'{\i}sica Te\'orica, Universidade Estadual Paulista, S\~ao Paulo, Brazil}
\affiliation{Simon Fraser University, Vancouver, British Columbia, and York University, Toronto, Ontario, Canada}
\affiliation{University of Science and Technology of China, Hefei, People's Republic of China}
\affiliation{Universidad de los Andes, Bogot\'{a}, Colombia}
\affiliation{Charles University, Faculty of Mathematics and Physics, Center for Particle Physics, Prague, Czech Republic}
\affiliation{Czech Technical University in Prague, Prague, Czech Republic}
\affiliation{Center for Particle Physics, Institute of Physics, Academy of Sciences of the Czech Republic, Prague, Czech Republic}
\affiliation{Universidad San Francisco de Quito, Quito, Ecuador}
\affiliation{LPC, Universit\'e Blaise Pascal, CNRS/IN2P3, Clermont, France}
\affiliation{LPSC, Universit\'e Joseph Fourier Grenoble 1, CNRS/IN2P3, Institut National Polytechnique de Grenoble, Grenoble, France}
\affiliation{CPPM, Aix-Marseille Universit\'e, CNRS/IN2P3, Marseille, France}
\affiliation{LAL, Universit\'e Paris-Sud, CNRS/IN2P3, Orsay, France}
\affiliation{LPNHE, Universit\'es Paris VI and VII, CNRS/IN2P3, Paris, France}
\affiliation{CEA, Irfu, SPP, Saclay, France}
\affiliation{IPHC, Universit\'e de Strasbourg, CNRS/IN2P3, Strasbourg, France}
\affiliation{IPNL, Universit\'e Lyon 1, CNRS/IN2P3, Villeurbanne, France and Universit\'e de Lyon, Lyon, France}
\affiliation{III. Physikalisches Institut A, RWTH Aachen University, Aachen, Germany}
\affiliation{Physikalisches Institut, Universit{\"a}t Freiburg, Freiburg, Germany}
\affiliation{II. Physikalisches Institut, Georg-August-Universit{\"a}t G\"ottingen, G\"ottingen, Germany}
\affiliation{Institut f{\"u}r Physik, Universit{\"a}t Mainz, Mainz, Germany}
\affiliation{Ludwig-Maximilians-Universit{\"a}t M{\"u}nchen, M{\"u}nchen, Germany}
\affiliation{Fachbereich Physik, Bergische Universit{\"a}t Wuppertal, Wuppertal, Germany}
\affiliation{Panjab University, Chandigarh, India}
\affiliation{Delhi University, Delhi, India}
\affiliation{Tata Institute of Fundamental Research, Mumbai, India}
\affiliation{University College Dublin, Dublin, Ireland}
\affiliation{Korea Detector Laboratory, Korea University, Seoul, Korea}
\affiliation{CINVESTAV, Mexico City, Mexico}
\affiliation{FOM-Institute NIKHEF and University of Amsterdam/NIKHEF, Amsterdam, The Netherlands}
\affiliation{Radboud University Nijmegen/NIKHEF, Nijmegen, The Netherlands}
\affiliation{Joint Institute for Nuclear Research, Dubna, Russia}
\affiliation{Institute for Theoretical and Experimental Physics, Moscow, Russia}
\affiliation{Moscow State University, Moscow, Russia}
\affiliation{Institute for High Energy Physics, Protvino, Russia}
\affiliation{Petersburg Nuclear Physics Institute, St. Petersburg, Russia}
\affiliation{Instituci\'{o} Catalana de Recerca i Estudis Avan\c{c}ats (ICREA) and Institut de F\'{i}sica d'Altes Energies (IFAE), Barcelona, Spain}
\affiliation{Stockholm University, Stockholm and Uppsala University, Uppsala, Sweden}
\affiliation{Lancaster University, Lancaster LA1 4YB, United Kingdom}
\affiliation{Imperial College London, London SW7 2AZ, United Kingdom}
\affiliation{The University of Manchester, Manchester M13 9PL, United Kingdom}
\affiliation{University of Arizona, Tucson, Arizona 85721, USA}
\affiliation{University of California Riverside, Riverside, California 92521, USA}
\affiliation{Florida State University, Tallahassee, Florida 32306, USA}
\affiliation{Fermi National Accelerator Laboratory, Batavia, Illinois 60510, USA}
\affiliation{University of Illinois at Chicago, Chicago, Illinois 60607, USA}
\affiliation{Northern Illinois University, DeKalb, Illinois 60115, USA}
\affiliation{Northwestern University, Evanston, Illinois 60208, USA}
\affiliation{Indiana University, Bloomington, Indiana 47405, USA}
\affiliation{Purdue University Calumet, Hammond, Indiana 46323, USA}
\affiliation{University of Notre Dame, Notre Dame, Indiana 46556, USA}
\affiliation{Iowa State University, Ames, Iowa 50011, USA}
\affiliation{University of Kansas, Lawrence, Kansas 66045, USA}
\affiliation{Kansas State University, Manhattan, Kansas 66506, USA}
\affiliation{Louisiana Tech University, Ruston, Louisiana 71272, USA}
\affiliation{Boston University, Boston, Massachusetts 02215, USA}
\affiliation{Northeastern University, Boston, Massachusetts 02115, USA}
\affiliation{University of Michigan, Ann Arbor, Michigan 48109, USA}
\affiliation{Michigan State University, East Lansing, Michigan 48824, USA}
\affiliation{University of Mississippi, University, Mississippi 38677, USA}
\affiliation{University of Nebraska, Lincoln, Nebraska 68588, USA}
\affiliation{Rutgers University, Piscataway, New Jersey 08855, USA}
\affiliation{Princeton University, Princeton, New Jersey 08544, USA}
\affiliation{State University of New York, Buffalo, New York 14260, USA}
\affiliation{Columbia University, New York, New York 10027, USA}
\affiliation{University of Rochester, Rochester, New York 14627, USA}
\affiliation{State University of New York, Stony Brook, New York 11794, USA}
\affiliation{Brookhaven National Laboratory, Upton, New York 11973, USA}
\affiliation{Langston University, Langston, Oklahoma 73050, USA}
\affiliation{University of Oklahoma, Norman, Oklahoma 73019, USA}
\affiliation{Oklahoma State University, Stillwater, Oklahoma 74078, USA}
\affiliation{Brown University, Providence, Rhode Island 02912, USA}
\affiliation{University of Texas, Arlington, Texas 76019, USA}
\affiliation{Southern Methodist University, Dallas, Texas 75275, USA}
\affiliation{Rice University, Houston, Texas 77005, USA}
\affiliation{University of Virginia, Charlottesville, Virginia 22901, USA}
\affiliation{University of Washington, Seattle, Washington 98195, USA}
\author{V.M.~Abazov} \affiliation{Joint Institute for Nuclear Research, Dubna, Russia}
\author{B.~Abbott} \affiliation{University of Oklahoma, Norman, Oklahoma 73019, USA}
\author{B.S.~Acharya} \affiliation{Tata Institute of Fundamental Research, Mumbai, India}
\author{M.~Adams} \affiliation{University of Illinois at Chicago, Chicago, Illinois 60607, USA}
\author{T.~Adams} \affiliation{Florida State University, Tallahassee, Florida 32306, USA}
\author{G.D.~Alexeev} \affiliation{Joint Institute for Nuclear Research, Dubna, Russia}
\author{G.~Alkhazov} \affiliation{Petersburg Nuclear Physics Institute, St. Petersburg, Russia}
\author{A.~Alton$^{a}$} \affiliation{University of Michigan, Ann Arbor, Michigan 48109, USA}
\author{G.~Alverson} \affiliation{Northeastern University, Boston, Massachusetts 02115, USA}
\author{G.A.~Alves} \affiliation{LAFEX, Centro Brasileiro de Pesquisas F{\'\i}sicas, Rio de Janeiro, Brazil}
% ...
%
%
\collaboration{The D0 Collaboration\footnote{with visitors from
$^{a}$Augustana College, Sioux Falls, SD, USA,
$^{b}$The University of Liverpool, Liverpool, UK,
$^{c}$SLAC, Menlo Park, CA, USA,
$^{d}$University College London, London, UK,
$^{e}$Centro de Investigacion en Computacion - IPN, Mexico City, Mexico,
$^{f}$ECFM, Universidad Autonoma de Sinaloa, Culiac\'an, Mexico,
and 
$^{g}$Universit{\"a}t Bern, Bern, Switzerland.
}} \noaffiliation
\vskip 0.25cm
    
\date{April 26, 2011}

\begin{abstract} 
We determine the fraction of \ttbar\ events with spin
correlation, assuming that the spin of the top quark is either
correlated with the spin of the anti-top quark as predicted by the standard model or is
uncorrelated. For the first time we use a matrix-element-based approach
to study  \ttbar\ spin correlation.
We use  \ttfull\ final states produced in
\ppbar\ collisions at a center of mass energy $\sqrt{s}=1.96$~TeV,
where $\ell$ denotes an electron or a muon. The data correspond to
an integrated luminosity of $5.4\text{~fb}^{-1}$ and were
collected with the \dzero\ detector at the Fermilab Tevatron
collider. 
The result agrees with the standard model prediction. We exclude 
the hypothesis that the spins of the \ttbar\ are uncorrelated at the $97.7\%$~C.L. 

\end{abstract}


\maketitle

While top and anti-top quarks are unpolarized in \ttbar\
production at hadron colliders and their spins cannot be
measured directly, their spins are correlated and this
correlation can be investigated experimentally~\cite{Barger:1988jj}. 
The standard model (SM) of particle physics predicts that
top quarks decay before fragmentation ~\cite{Bigi:1986jk}, which is in
agreement with the measured lifetime of the top quark~\cite{d0_width}.
The information on the spin orientation of top quarks is
transferred through weak interaction  to the angular distributions of the
decay products ~\cite{spin_theory,Brandenburg:2002xr}.

We present a test of the hypothesis that the correlation of the spin of $t$ 
and $\bar{t}$ quarks is as expected in the SM as opposed to the
hypothesis that they are uncorrelated. The spins could become
decorrelated if the spins of the top quarks flip before they decay or
if the polarization information is not propagated to all the final
state products. This could occur if the top quark
decayed into a scalar charged Higgs boson and a $b$ quark ($t \to H^+
b$)~\cite{ratio,HplusD0,HplusCDF}.
 
Recently, the CDF Collaboration has presented a measurement of the \ttb\
spin correlation parameter $C$  
in semileptonic final states from a differential
angular distribution~\cite{cdf_spin}. The spin correlation strength
$C$ is defined by  $d^{2}\sigma / d\cos\theta_1 d\cos\theta_2 =\sigma
( 1 - C \cos\theta_1 \cos\theta_2 )/4 $,
where $\sigma$ denotes the cross section, and $\theta_1$ and
$\theta_2$ are the angles between the direction of flight of the decay
leptons (for leptonically decaying $W$ bosons) or jets (for
hadronically decaying $W$ bosons) in the parent $t$ and $\bar{t}$ rest
frames and the spin quantization axis. 
The value $C=+1$ ($-1$) gives fully
correlated (anticorrelated) spins and $C=0$ corresponds to no spin
correlation, while the NLO SM prediction using the beam momentum
vector as spin 
quantization axis is $C=0.777^{+0.027}_{-0.042}$~\cite{spin_theory}. 
The D0 Collaboration has performed two measurements of $C$  
in dilepton final states~\cite{D0RunI,d0_spin}, where the second analysis 
uses the same dataset as this measurement. None of the previous analyses has
sufficient sensitivity to distinguish between a hypothesis of
no correlation and of correlation as predicted by the SM. 

In this Letter, we present the first measurement of  spin correlation
in \ttbar\ production 
using a matrix-element-based approach, exploring the full
matrix elements (ME) in leading order (LO) Quantum Chromodynamics (QCD). 
We extract 
the fraction $f$ of \ttbar\ candidate
events where the \ttbar\ spin correlation is as predicted by the SM
over the total number of \ttbar\ candidate events assuming that
they consist of events with SM spin correlation and of events without
spin correlation. We use  \ttbar\ event candidates with two
charged leptons in the final state, where the charged leptons
correspond to either electrons or muons, in a dataset of 5.4~fb$^{-1}$
of integrated luminosity that has been collected with the D0 
detector at the Fermilab Tevatron $p\bar{p}$ collider.  
With a matrix-element-based approach, we use the full kinematics  of 
the final state to improve the sensitivity with respect to using only 
a single distribution by almost 30\%.

The D0 detector~\cite{d0det} comprises  a tracking system, a
calorimeter, and a muon spectrometer. The tracking system consists of a silicon
microstrip tracker  and a central fiber tracker, both
located inside a 2~T superconducting solenoid.  The system provides
efficient charged-particle tracking in the pseudorapidity region
$|\eta_{\text{det}}| < 3$~\cite{eta}.  The calorimeter has a
central section  covering $|\eta_{\text{det}}|<1.1$ and two end
calorimeters (EC) extending coverage to $|\eta_{\text{det}}|\approx 4.2$ for jets.
The muon system surrounds the calorimeter and consists of three layers
of tracking detectors and scintillators covering
$|\eta_{\text{det}}|<2$~\cite{muon_detector}.  A 1.8~T toroidal iron magnet
is located outside the innermost layer of the muon detector.  The
integrated luminosity is calculated from the rate of inelastic {\mbox{$p\bar p$}}\
collisions, measured with plastic scintillator arrays that
are located in front of the EC.

We use the same selection of $\ell\ell$ ($ee$, $e\mu$, and $\mu\mu$)
events as  described in Ref.~\cite{d0_spin}, therefore only a short overview of the selection  is given. 
To enrich the data sample in \ttbar\ events, we require two isolated, oppositely
charged leptons with $p_T>15$~GeV and at least two
jets with $p_T>20$~GeV and $|\eta_{\text{det}}|<2.5$.  Electrons in the central~($|\eta_{\text{det}}|<1.1$) and forward~($1.5<|\eta_{\text{det}}|<2.5$) region 
are accepted, while muons must satisfy $|\eta_{\text{det}}|<2$. Jets are reconstructed with a mid-point cone algorithm~\cite{jetcone}
with radius ${\cal R}=0.5$. Jet energies are corrected for calorimeter
response, additional energy from noise, pileup, and multiple \ppbar\
interactions in the same bunch crossing, and out-of-cone shower
development in the calorimeter. We require three or more tracks originating from the
selected \ppbar interaction vertex within each jet cone.  The high
instantaneous luminosity achieved by the Tevatron leads to a
significant background contribution from additional \ppbar\ collisions within the
same bunch crossing. The track requirement removes jets from such additional 
collisions and is only necessary for data taken after the initial
1~fb$^{-1}$. The missing transverse energy (\met) is defined by the magnitude of the  negative
 vector sum of all 
transverse energies measured in calorimeter cells, corrected for the
transverse energy of isolated muons and for the different response  to electrons and jets. 
A more detailed description of objects reconstruction can
be found in~\cite{Abachi:2007:prd}.

The final selection in the $e\mu$ channel requires that the 
scalar sum of the leading lepton $p_T$ and the $p_T$ of the two most
energetic jets be greater than 110~GeV.  
To reject background in $ee$ and $\mu\mu$ events, where \met\ arises
from mismeasurement,  
we compute a \met\ significance 
which takes into account the resolution of the lepton and jet measurements. 
We require the significance to exceed five standard deviations.  
In the $\mu\mu$ channel, events are furthermore required to have $\met > 40$~GeV.

The \ttbar\ signal is modeled using the \mcatnlo~\cite{mcatnlo}
event generator together with the CTEQ6M1 parton distribution function (PDF)
\cite{bib:CTEQPDF}, assuming a top quark mass $m_{t}=172.5$~GeV. 
We generate \ttb\ Monte Carlo (MC)
samples with and without the expected spin correlation, as both
options are available in \mcatnlo. The events are processed through
\herwig~\cite{herwig} to simulate fragmentation, hadronization and
decays of short-lived particles and through a full detector simulation
using \geant~\cite{geant}. We overlay data events
from a random bunch crossing to model the effects of detector noise
and additional \ppbar\ interactions to the MC events. The same 
reconstruction programs are used to process the data and MC simulated events.

Sources of background arise from the production of electroweak bosons
that decay into charged leptons. In the $ee$, $e\mu$, and $\mu\mu$
channels, the dominant backgrounds are Drell-Yan processes, namely $Z/\gamma^{\ast}\rightarrow e^+e^-$, 
$Z/\gamma^{\ast}
\rightarrow \tau^+\tau^- \rightarrow \bar{\nu}
\ell^+ \nu \nu \ell^- \bar{\nu}$, with
$\ell^\pm=e^\pm$~or~$\mu^\pm$, and $Z/\gamma^{\ast} \rightarrow
\mu^+\mu^-$. 
In addition,  diboson production  
($WW$, $WZ$ and $ZZ$) contributes when the bosons decay to two charged leptons. We model the $Z/\gamma^{\ast}$ background with  \alpgen~\cite{alpgen}, interfaced with
\pythia~\cite{pythia},  while diboson production is simulated using \pythia\ only.
The $Z/\gamma^{\ast}$ and diboson
processes are generated at LO and are normalized to
the next-to-next-to-leading order (NNLO) inclusive cross~section for
$Z/\gamma^{\ast}$ events and to the next-to-leading order (NLO)
inclusive cross~sections for diboson
events~\cite{campbell-ellis,mcfm}. For all background processes the 
CTEQ6L1 PDF~\cite{bib:CTEQPDF}) are used.  

 Detector-related backgrounds can be attributed to jets
mimicking electrons, muons from semileptonic decays of $b$ quarks,  in-flight decays of pions or kaons in a jet, and misreconstructed 
\met. These backgrounds are modeled with data. Background from electrons that arise from
jets comprising an energetic $\pi^0$ or $\eta$ particle and an
overlapping track is estimated from the distribution
of an electron-likelihood discriminant in data~\cite{Abachi:2007:prd}. In the $e\mu$ and
$\mu\mu$ channels, muons produced in jets that fail to be
reconstructed can appear isolated. 
Table~\ref{tab:yields} summarizes the yields for the signal and background contributions. 

\begin{table}
\begin{center}
\caption{\label{tab:yields}Yields of selected events. The number of \ttbar\ events is 
calculated using the measured cross section of $\sigmatt=8.3$~pb and the 
measured $f=0.74$. Uncertainties include statistical and systematic contributions. }
\begin{tabular}{cccccc}
\hline
\hline
\ttbar\ & $Z/\gamma^{*}$ & Diboson & Instrumental & Total & Observed\\
$341\pm 30$ & $93\pm 15$ & $19 \pm 3$  & $28 \pm 5$ & $481 \pm 39$ & 485\\
\hline
\hline
\end{tabular}
\end{center}
\end{table}

\begin{figure}[htbp]
\begin{center}
\includegraphics[width=0.45\textwidth]{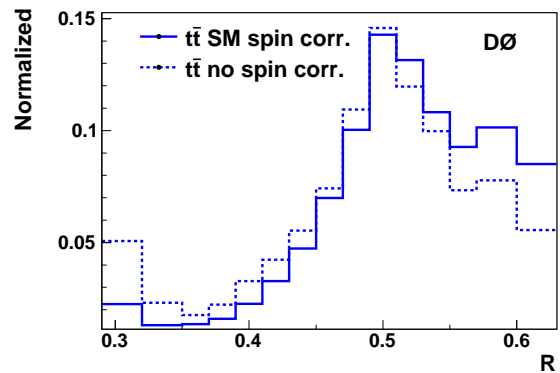}
\end{center}
\caption{Comparison of the discriminant $R$ between SM spin correlation
$H=c$  and no spin correlation $H=u$  at parton level. 
The first and last bin include also  the contributions
from $R<0.29$ and $R>0.63$.}
\label{fig:D_parton}
\end{figure}

To distinguish the hypothesis $H$ of correlated top quark spins 
as predicted by the SM ($H=c$) from the hypothesis of uncorrelated top quark 
spins ($H=u$), we calculate a discriminant $R$~\cite{melnischulze}
defined as
\begin{equation}
R = \frac{P_{\rm sgn} (H=c)}{P_{\rm sgn}(H=u) + P_{\rm sgn} (H=c)}\,,
\label{eq:discr}
\end{equation}
where we calculate per-event probability densities, $P_{\rm sgn}$, for
\ttbar\ signal events for both hypotheses constructed from the
LO MEs ${\cal M}(y, H)$~\cite{Mahlon_Parke},
\begin{eqnarray}
\label{eq:psgn}  \nonumber
P_{\rm sgn}(x;H) & & \hspace{-0.3cm}= \frac{1}{\sigma_{\rm obs}} \int
~f_{\rm PDF}(q_1)~f_{\rm PDF}(q_2) 
{\rm d}q_1 {\rm d}q_2 \\ 
&\cdot & \frac{(2 \pi)^{4} \left|{\cal M}(y,H)\right|^{2}}
       {q_1q_2 s}
  ~ W(x,y) 
~     {\rm d}\Phi_{6}.
\end{eqnarray}
Here,  $\sigma_{\rm obs}$ denotes the leading order cross section
including selection efficiency, 
$q_1$ and $q_2$ the energy fraction of the
incoming quarks from the proton and antiproton, respectively, $f_{\rm PDF}$ the
parton distribution function, $s$ the center-of-mass
energy squared and ${\rm d}\Phi_{6}$
the infinitesimal volume element of the 6-body phase space. 
The detector resolution is taken into account through a transfer
function $W(x,y)$ that describes the probability of a partonic final
state $y$ to be measured as $x = (\tilde{p}_1,\dots,\tilde{p}_n)$,
where $\tilde{p}_i$ denotes the measured four-momenta of the final
state particles.
For hypothesis $H=c$ we use the ME for the full process $q \bar{q} \to
\ttbar\ \to W^+ b \, W^- \bar{b} \to \ell^+ \nu_{\ell} b \, \ell'^-
\bar{\nu}_{\ell'} \bar{b}$ averaged over the initial quarks' color
and spin and 
summed over the final colors and spins~\cite{Mahlon_Parke}. For hypothesis
$H=u$, we use the ME of the same process neglecting the spin correlation
between production and decay~\cite{Mahlon_Parke}. 
The \ttbar\ production cross section, $\sigmatt$, does not depend on the hypothesis $H=c$
or $H=u$, and  is taken as identical for both hypotheses.
It is assumed that momentum directions for jets 
and charged leptons and the electron energy are well measured, leading
to a reduction of the number of integration dimensions. Furthermore,
the known masses of the final state particles are used as input, and
it is assumed that the \ttbar\ system has no transverse momentum
resulting in a six dimensional phase space integration. 
More details of the calculation of $P_{\rm sgn}$ can
be found in~\cite{menimpaper}.
Figure~\ref{fig:D_parton} shows the discriminant $R$ for generated
partons 
for $H=c$ and $H=u$ for \ttbar\ MC events.

To measure the fraction $f_{\mathrm{meas}}$ of events with SM spin correlation,
we build templates of $R$ distributions for signal MC with and 
without spin correlation as well as for each source of background.  
The templates are compared to the $R$ distribution in data and the 
fraction of events with SM spin correlation  is extracted.

\begin{figure}[h]
\begin{centering}
\includegraphics[width=0.45\textwidth]{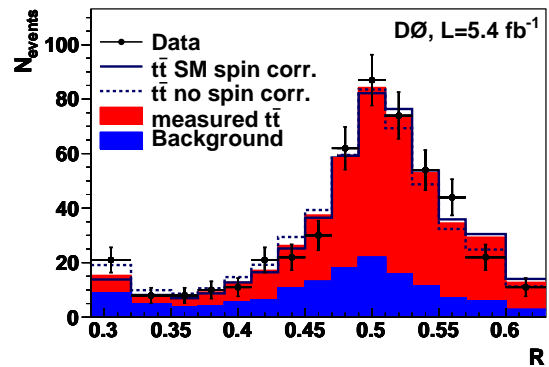}
\par\end{centering}
\caption{\label{fig:D_template}
(Color online) The predicted discriminant distribution $R$ for
the combined dilepton event sample for the fitted  $\sigmatt$ 
and  $f_{\mathrm{meas}}$ compared to the data.
The  prediction with spin correlation ($f=1$) and without spin
correlation ($f=0$) is shown including background.
The first and last bin include also  the contributions
from $R<0.29$ and $R>0.63$. }

\end{figure}

In Fig.~\ref{fig:D_template}, the measured discriminant $R$ in data
is compared to templates for \ttbar\ production with SM spin correlation  and without 
spin correlation including background  for all dilepton channels combined. 
The separation  between $H=c$ and $H=u$ is decreased compared to the parton level.

We perform a binned maximum likelihood fit to the $R$
distribution to extract $f_{\mathrm{meas}}$  by
fitting 
\begin{equation}
\label{eq:fsm}
 m^{(i)}  =  f _{\mathrm{meas}}\, m_{\rm c}^{(i)} + (1-f_{\mathrm{meas}}) \, m_u^{(i)} + \sum_j m_j^{(i)} \;, 
\end{equation}
where $m_{\rm c}^{(i)}$ is the predicted
number of events in bin $i$ for 
the signal template including SM spin correlation, $m_u^{(i)}$ is the
predicted number of events in bin $i$ for the template without spin
correlation and $ \sum_j m_j^{(i)}$ is the sum over all
background contributions $j$ in bin $i$. 
To remove the dependence on the absolute
normalization, we calculate the predicted number
of events, $m^{(i)}$, as a function of $f_{\mathrm{meas}}$ and $\sigmatt$ and extract both simultaneously.

The likelihood function
\begin{equation}
{\cal L} = \prod_{i}^{N} {\cal P}(n^{(i)}, m^{(i)})
\times \prod_{k=1}^{K} {\cal G}(\nu_k;0,{\text{SD}}_k) \,,
\label{eq:mlikeli}
\end{equation}
is maximized with ${\cal P}(n,m)$ representing the Poisson probability
to observe $n$ events when $m$ events are expected. The first product
runs over all  bins $i$ of the templates in all
channels. Systematic uncertainties are taken into account by
 parameters $\nu_k$, where each independent source of
systematic uncertainty $k$ is modeled as a Gaussian probability
density function, ${\cal G}\left(\nu;0,{\text{ SD}}\right)$, with zero
mean and an rms corresponding to one standard deviation (SD) in the
uncertainty of that parameter. Correlations among systematic
uncertainties between channels are taken into account by using a
single parameter for the same source of uncertainty. 

We distinguish between systematic uncertainties that only affect the 
yield of signal or background, and those that change the shape of the $R$ distribution. 
We consider the jet energy scale, jet energy resolution, jet 
identification, PDFs, background modeling, and the choice
of $m_t$ in the calculation of $P_{\rm sgn}$ as uncertainties affecting the shape of $R$. 
Systematic uncertainties on normalizations include lepton identification,
trigger requirements, uncertainties on the normalization of
background, the uncertainty on the luminosity, MC modeling, and the
determination of instrumental background. We also include an
uncertainty on the templates because of limited statistics in the MC
samples. 

The statistical and systematic uncertainties on 
$f_{\mathrm{meas}}$ are given in Table~\ref{tab:systematics_f}.  We evaluate
the size of the individual sources of systematic uncertainty by calculating 
$f_{\mathrm{meas}}$ and $\sigmatt$ using the  parameters $\nu_k$ shifted by $\pm1$SD from their fitted mean.

\begin{table}
\begin{center}
\caption{\label{tab:systematics_f} Summary of uncertainties on
  $f_{\mathrm{meas}}$. 
} 
\begin{tabular}{ccc} \hline \hline
 Source  &        +1SD    &   $-1$SD   \\ \hline
                 Muon identification & 0.01 & -0.01 \\ 
              Electron identification and smearing & 0.02 & -0.02 \\ 
                                   PDF & 0.06 & -0.05 \\ 
                                       $m_t$ & 0.04 & -0.06 \\ 
                                          Triggers & 0.02 & -0.02 \\ 
                       Opposite charge selection & 0.01 & -0.01 \\ 
                                       Jet energy scale & 0.01 & -0.04 \\ 
            Jet reconstruction and identification & 0.02 & -0.06 \\ 
                                     Background normalization & 0.07 & -0.08 \\ 
                            MC statistics & 0.03 & -0.03 \\ 
                    Instrumental background & 0.01 & -0.01 \\ 
                                    Integrated luminosity & 0.04 & -0.04 \\ 
                                             Other & 0.02 & -0.02 \\ 
            MC statistics for template fits & 0.10 & -0.10 \\ 
                                 \hline  
                    Total systematic uncertainty & 0.15 & -0.18 \\ 
\hline  
  Statistical uncertainty &       0.33 &   -0.35  \\ \hline \hline
 \end{tabular}
 \end{center}
\end{table}

\begin{figure}[h]
\begin{centering}
\includegraphics[width=0.5\textwidth]{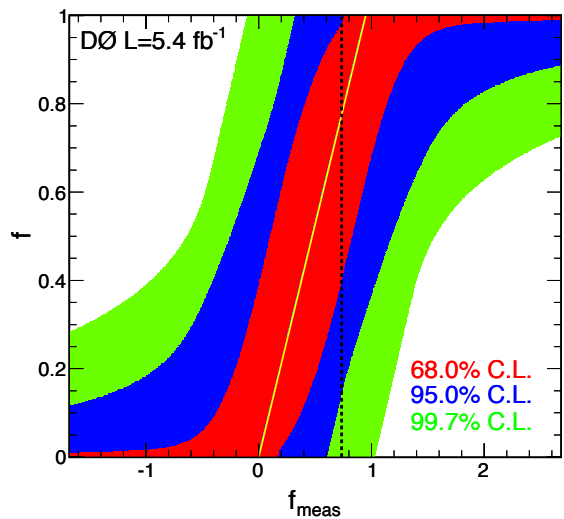}
\par\end{centering}

\caption{\label{fig:fc_f}
(Color online) For all channels the 68.0\% (inner), 95.0\% (central), and
99.7\% (outer) C.L. bands of $f$ as a function of $f_{\mathrm{meas}}$
from likelihood fits to MC events.  The thin yellow line indicates the
most probable value of $f$ as a function of $f_{\mathrm{meas}}$, and
therefore represents the calibration of the method.  The vertical
dashed black line indicates the measured value
$f_{\mathrm{meas}}=0.74$.}
\end{figure}

To estimate the expected uncertainty on the result, ensembles of MC
experiments are generated for different values of 
$f$, and the maximum likelihood fit is repeated, yielding
a distribution of $f_{\mathrm{meas}}$ for each generated $f$.
Systematic uncertainties are included in this procedure, taking
correlations between channels into account. We then apply the ``ordering
principle'' for ratios of likelihoods~\cite{fc_limit} to the
distributions of  $f_{\mathrm{meas}}$ and generated
$f$, without constraining $f_{\mathrm{meas}}$ to
physical values. The resulting allowed regions for different
confidence levels as a function of $f_{\mathrm{meas}}$ and $f$ are
shown in Fig.~\ref{fig:fc_f}. 
From the maximum likelihood fit to
data, we obtain
\begin{equation}
f_{\mathrm{meas}} = 0.74^{+0.40}_{-0.41} \, ({\text{stat+syst}}) \,\,.
\end{equation}
The simultaneously extracted \ttbar\ cross section is found to be
\begin{equation}
\sigmatt = 8.3 ^{+1.1}_{-0.9} \, ({\text{stat+syst}})  \ \, {\text{pb}} 
\end{equation}
for $m_t=172.5$~GeV and in good agreement with the SM prediction
of $\sigmatt = 7.46 ^{+0.48}_{-0.67} \ \,
{\mathrm{pb}}$~\cite{SMtheory_MochUwer}. 
The comparison of $f$ for prediction and data with the fitted result
is shown in Fig.~\ref{fig:D_template}.
The measured fraction is consistent with the
SM expectation ($f=1$) and we exclude the no-correlation
hypothesis ($f=0$) at the 97.7\% C.L.
For the SM value of $f=1$ we expect to exclude the hypothesis $f=0$ with $99.6\%$~C.L.

Assuming $f_{\mathrm{meas}}$ and using the full matrix elements
for \ttbar\ production with SM spin correlation or without spin
correlation, other observables can be extracted to study the impact of this
measurement. For illustration, we derive $C$ from the measured value of 
$f$ and the NLO prediction of $C$ in the SM, yielding  $C_{\mathrm{meas}}=0.57
\pm 0.31 \,({\text{stat+syst}})$~\footnote{This value is derived by
multiplying $f_{\mathrm{meas}}$ with the NLO SM prediction of
$C=0.777^{+0.027}_{-0.042}$~\cite{spin_theory}. 
The physical region of this
measurement is $0 \le f_{\mathrm{meas}} \le 1$ which translates into a
requirement of $0 \le C_{\mathrm{meas}} \le 0.777$. This is in
contrast to the measurement of $C_{\mathrm{meas}} =
0.10^{+0.42}_{-0.44} \, ({\text{stat+syst}}) $ 
in~\cite{d0_spin} where $-1 \le C_{\mathrm{meas}} \le 1$ is
allowed. Therefore, both results have to be compared with caution.}.

In summary, we have presented the first measurement of the fraction of
$t\bar{t}$ events with correlated spins using a matrix element technique.
This fraction can be translated into the most precise value of the
correlation strength $C_{\mathrm{meas}}$ to date.

We wish to thank W.~Bernreuther, K. Melnikov, S.~J.~Parke, and M. Schulze
for fruitful discussions regarding this analysis.
We thank the staffs at Fermilab and collaborating institutions,
and acknowledge support from the
DOE and NSF (USA);
CEA and CNRS/IN2P3 (France);
FASI, Rosatom and RFBR (Russia);
CNPq, FAPERJ, FAPESP and FUNDUNESP (Brazil);
DAE and DST (India);
Colciencias (Colombia);
CONACyT (Mexico);
KRF and KOSEF (Korea);
CONICET and UBACyT (Argentina);
FOM (The Netherlands);
STFC and the Royal Society (United Kingdom);
MSMT and GACR (Czech Republic);
CRC Program and NSERC (Canada);
BMBF and DFG (Germany);
SFI (Ireland);
The Swedish Research Council (Sweden);
and
CAS and CNSF (China).


\begin{thebibliography}{99}
\bibitem{eta} The pseudorapidity $\eta$ of a particle is defined as function
    of the polar angle $\theta$ as
    $\eta(\theta) = - \ln[\tan(\theta/2)]$.  We use here  detector $\eta$
    ($\eta_{\text{det}}$) which  is defined
    with respect to the center of the detector.
\end{thebibliography}
\end{document}